\documentclass[12pt]{article}
\def\cF{{\cal F}}

\newfont{\goth}{eufm10 scaled \magstep1}

\def\a{\alpha}

\def\b{\beta}

\def\c{\gamma}
\def\d{\delta}
\def\e{\epsilon}

\def\s{\sigma}
\def\t{\tau}

\def\beq{\begin{equation}}\def\eeq{\end{equation}}
\def\beqa{\begin{eqnarray}}\def\eeqa{\end{eqnarray}}
\def\barr{\begin{array}}\def\earr{\end{array}}

\def\del{\partial}

\def \ys {{y\kern-.5em / \kern.3em}}

\let\bm=\bibitem

\def\nn{\nonumber}
\def\bd{\begin{document}}
\def\ed{\end{document}}
\def\ba{\begin{array}}
\def\ea{\end{array}}
\def\bea{\begin{eqnarray}}
\def\eea{\end{eqnarray}}
\def\ft#1#2{{\textstyle{{\scriptstyle #1}\over {\scriptstyle #2}}}}
\def\fft#1#2{{#1 \over #2}}
\newcommand{\be}{\begin{equation}}
\newcommand{\ee}{\end{equation}}
\newcommand{\eq}[1]{(\ref{#1})}
\def\eqs#1#2{(\ref{#1}-\ref{#2})}
\def\det{{\rm det\,}}
\def\tr{{\rm tr}}
\newcommand{\ho}[1]{$\, ^{#1}$}
\newcommand{\hoch}[1]{$\, ^{#1}$}
\def\ra{\rightarrow}
\def\uha{{\hat {\underline{\a}} }}
\def\uhc{{\hat {\underline{\c}} }}

\def \Om {\Omega}
\def \bfd {{\bf d}}
\def \del {\partial}
\def \eps {\epsilon}
\def \Z {{\bf Z}}
\def \xb {\bar{x}}
\def \la {\langle}
\def \ra {\rangle}
\def \Omt {\tilde \Omega} 
\def \la {\langle}
\def \ra {\rangle}

\def \II {I\hspace{-.1em}I\hspace{.2em}}

\begin{document}

\hfill{NEIP-00-002}

\hfill{hep-th/0001144}

\vspace{20pt}

\begin{center}

{\Large \bf Noncommutative Open String: Neutral and Charged}
\footnote{Talk presented at the  
Anttila winter school and workshop on
``String theory and Gauge theory''  (Anttila, Sweden, Feb 1999), 
and the 
International Workshop ``Supersymmetries and Quantum Symmetries'' 
(Dubna, Russia, July 1999)
}

\vspace{30pt}

{\large Chong-Sun Chu}

\vspace{15pt}

{\em Institute of Physics, University of Neuch\^atel, 
CH-2000 Neuch\^atel, Switzerland}

\vskip .2in \sffamily{chong-sun.chu@iph.unine.ch} 

\vspace{60pt}

{\bf Abstract}

\end{center}

We review the quantization of open string in 
NS-NS background and demonstrate that 
its endpoint becomes noncommutative. 
The same approach allows us to determine the
noncommutativity that arises  for a charged open 
string in background gauge fields. While NS-NS background is relevant
for ``worldvolume'' noncommutativity, a  simple argument suggests that 
RR background is likely  
to be relevant for ``spacetime'' noncommutativity.

\newpage


\section{Introduction}

String theory has experienced remarkable progress in the last couple
of years. One of the recent interests is the realization that
noncommutative spacetime arises naturally in string and M-theory. 
The Matrix theory proposal  \cite{BFSS} conjecture
that M theory can be defined by a supersymmetric quantum mechanics.
Upon compactification,  Matrix theory  is described by a
supersymmetric Yang-Mills  living on the dual torus 
\cite{BFSS,taylor}. The situation is however more complicated when
there is a background field. It was proposed  by  
Connes, Douglas and Schwarz \cite{CDS} that
Matrix model compactified on a $T^2$ give rises to
noncommutative SYM when there is a background field $C_{-12}$. Since
Matrix model can be obtained by discretizing the supermembrane, one
approach to obtain the Matrix model with background $C$-field is to
discretize the supermembrane theory with a WZ coupling term \cite{CHL}. 
It turns out
the resulting Matrix model is related to the original Matrix model
without background field by a singular similarity transformation; and 
the Moyal product as well as the Seiberg-Witten map between
commutative and noncommutative variables  are correctly reproduced
upon compactification \cite{CHL}. 

From the string theory point of view, the above $C$-field background of
M-theory corresponds to string theory with  a NS-NS $B$-field
background; and the noncommutativity over the D-brane worldvolume 
can  be shown to arises from the open string point of view 
\cite{DH,m1,m2,HV,CH,CH1,Schom,SW}. 
Various aspects of
noncommutative geometry in string theory were further 
examined in \cite{SW}. See \cite{ncg1,ncg2} for a detail exposition
of noncommutative geometry, with motivations and applications in
physical problems. 

In the following, we will follow the Hamiltonian approach taken in
\cite{CH} for open string quantization 
in background $B$-field. This has the
advantage of being easily generalizable
to the case of a charged open string in 
background gauge fields.

\section{String Theory in Constant NS-NS Background}

Consider a fundamental string ending on a D$p$-brane in the presence
of a $B$-field. The bosonic part of the action takes the form
\be \label{action} S_B= {1 \over 4\pi\alpha'}
\int_{\Sigma} d^2\sigma \bigl[ g^{\a\b}G_{\mu\nu} \partial_\a
X^{\mu}\partial_\b X^{\nu}+ \eps^{\a\b} B_{\mu\nu}\partial_\a
X^{\mu}\partial_\b X^{\nu} ] + {1 \over 2\pi\alpha'}\int_{\partial
\Sigma} d \tau A_i(X) \partial_{\tau}X^i, 
\ee 
where $A_i,\ i=0,1,\cdots, p$, is the $U(1)$ gauge 
field living on the D$p$-brane and the string background is 
$ G_{\mu\nu} = \eta_{\mu\nu},\Phi =\mbox{constant},
H=dB=0 $. 
We use the convention $\eta^{\a\b}=\mbox{diag}(-1,1)$ and
$\eps^{01}=1$ as in \cite{CH}.  
This can be in type 0 superstring, type \II superstring, or in the bosonic
string theory. 
\footnote{
With slight modification, the considerations here 
can also be applied to study open string ending on a D-brane in
type I string theory. There a quantized $B$-field appears \cite{qB} 
and natively
one expect  that a noncommutative gauge theory with $SO$ or $SP$ gauge
group to appear. A more careful analysis shows however that the
resulting gauge theory 
is equivalent to one without deformation by
doing a field redefinition. It remains a challenge to find out how to
define noncommutative gauge theory with gauge group other than $U(N)$
\cite{SW} and in what setting of string theory they arise.
I am grateful to  Bogdan Morariu and Bruno Zumino 
for carrying out this analysis together. 
}
If both ends of the string are attached to the same D$p$-brane, the last
term in (\ref{action}) can be written as 
\be
\frac{-1}{4\pi\a'}\int_{\Sigma}d^2\s
\eps^{\a\b}F_{ij}\del_{\a}X^i\del_{\b}X^j.  
\ee 
Furthermore, consider
the case $B=\sum_{i,j=0}^{p}B_{ij}dX^i dX^j$, then the action
(\ref{action}) can be written as 
\be 
S_B=-\int d\t
L=\frac{1}{4\pi\a'}\int d^2\s \bigl[ g^{\a\b}\eta_{\mu\nu}\partial_\a
X^{\mu}\partial_\b X^{\nu}+ \eps^{\a\b}\cF_{ij}\partial_\a
X^{i}\partial_\b X^{j} ].  
\ee
Here 
\be \cF =B-dA=B-F 
\ee 
is the modified Born-Infeld field strength and $x_0^a$
is the location of the D-brane.  Indices are raised and lowered by
$\eta_{ij} = (-, +, \cdots, +)$.
 
One obtains the equations of motion 
\be \label{eom}
(\del^2_{\tau}-\del^2_{\s}) X^\mu =0
\ee 
and the boundary conditions at $\s =0, \pi$: 
\bea
&\del_\s X^i + \del_\tau X^j \cF_j{}^i =0, \quad i,j= 0,1,\cdots, p,
\label{BC1}\\
&X^a =x_0^a,\quad a =p+1, \cdots, D. \label{BC2} 
\eea
The mode expansion that solve \eq{BC1} is
\be
X^k =x_0^k+(p_0^k \tau -  p_0^j \cF_j{}^k \sigma)+ 
\sum_{n\neq 0} {e^{-in\tau} \over n}
\bigl(ia^k_n \cos n\sigma -   a_n^j \cF_j{}^k \sin n\sigma \bigr). 
\label{mode1}
\ee
This implies that the canonical momentum 
$ 2\pi\a' P^k(\t,\s)=\del_\tau X^k + \del_\sigma X^j\cF_j{}^k,$
has the expansion
\be
2\pi\a' P^k(\t,\s)=\{ p_0^l +
\sum_{n\neq 0} a_n^l e^{-in\tau} \cos n\sigma \}M_l{}^k, \label{mode2}
\ee
where  $M_{ij}=\eta_{ij}-\cF_i{}^k\cF_{kj}$.  

The constraint \eq{BC2} is standard.  We will be mainly 
interested in the
constraint \eq{BC1}.  As demonstrated in \cite{CH}, the BC \eq{BC1}
implies that 
\be \label{dXwithP} 
2\pi\a' P^k(\t,0)\cF_k{}^i =-\del_{\s}X^j(\t,0)M_j{}^i. 
\ee 
It follows that
\bea\label{PP1}
&2\pi\a'[P^k(\t,0),P^j(\t,\s')]\cF_k{}^i= 
-\del_{\s} [X^k(\t,\s),P^j(\t,\s')]_{\s=0} \, M_k{}^i, \\
& 2\pi\a'[P^k(\t,0),X^j(\t,\s')] \cF_k{}^i = 
- \del_\s [X^i(\t,\s), X^j(\t,\s')]_{\s=0}.  \label{XX1}
\eea
These simple relations show that the standard canonical commutation
relations for $\cF=0$,
\bea 
&[X^i(\t,\s),P_j(\t,\s')]=i\d^i_j\d(\s,\s'), \label{stdCR1} \\
&[P_i(\t,\s),P_j(\t,\s')]=0,\label{stdCR2} \\ 
&[X^i(\t,\s),X^j(\t,\s')]=0,  \label{stdCR3}
\eea 
are not 
compatible with the boundary condition \eq{BC1} when 
$\cF\neq 0$. Since the modified boundary condition \eq{BC1} occurs 
only at the boundary, it is clear that the commutation relations
\eq{stdCR1}-\eq{stdCR3} are modified only there. 

Following the procedure of \cite{nonab},
one finds that the symplectic form is 
\be\label{omega}
\Omega= \int_0^\pi d\s  \bfd P_{\mu} \bfd{X}^{\mu},
\ee
because the 
modifications to \eq{stdCR1}-\eq{stdCR3} occur on a measure zero
set and so do not modify the familiar form of $\Omega$. To determine
how the commutation relations are modified, we evaluate \eq{omega} for
the mode expansions \eq{mode1} and \eq{mode2}
to get the Poisson structure for the modes. To be
consistent, the resulting expression should be $\t$-independent.  
Using \eq{eom} and \eq{BC1}, 
it is easy to check that this is indeed the case.
Substituting the mode expansions \eq{mode1}, \eq{mode2}, one obtains
\be \label{modeF}
\Om=\frac{1}{2\a'}\left\{M_{ij}\bfd p_0^i
(\bfd x_0^j+\frac{\pi}{2}\cF^j{}_k\bfd p_0^k)+
\sum_{n > 0}\frac{-i}{n}(M_{ij}\bfd a^i_n\bfd a^i_{-n}
+\bfd a^a_n\bfd a^a_{-n})\right\},
\ee
which is explicitly time independent
\footnote{This corrects an  irrelevant step  in \cite{CH}.}.
Eqn. \eq{modeF}  
implies the following commutation relations for the modes
\bea
&[a_n^i, x_0^{j}]=[a_n^i, p_0^j] =0, \quad
[a_m^i, a_n^j]=2\a'mM^{-1ij}\d_{m+n}, \label{cr2} \\
&[p_0^i, p_0^j]=0, \quad
[x_0^i, p_0^j]=i2\a'M^{-1ij}, 
\quad [x_0^i,x_0^j]= i 2\pi\a'(M^{-1}\cF)^{ij},
\eea
and in turn implies
\bea
&[P^i(\t,\s),P^j(\t,\s')] =0, \label{F1} \\
&[X^k(\t,\s),X^l(\t,\s')] = \left\{
\begin{array}{ll}
\pm 2\pi i \a'( M^{-1} \cF )^{kl}, & \s=\s' =0 \mbox{ or } \pi,\cr 
0, &  \mbox{otherwise},  
\end{array}
\right. \label{F2} \\
&[X^i(\t,\s),P^j(\t,\s')]= i \eta^{ij} \d(\s,\s'), \label{F3}
\eea
where
$\d(\s,\s')$ is the delta function on $[0,\pi]$ with vanishing
derivative at the boundary, 
$ \d(\s,\s')= \frac{1}{\pi}
\left(1+ \sum_{n\neq 0}\cos n\s \cos n\s' \right). $
Thus we see that the string becomes noncommutative at the endpoint, 
i.e. the D-brane becomes noncommutative. Note that the 
noncommutativity depends on quantity defined on the
D-brane. The relation \eq{F2} is manifestly local. 

We finally remark that since the ghost system is not sensitive
to the presence of $\cF$; their boundary condition and hence their
central charge is not modified by  $\cF$. Therefore to be free from
conformal anomaly, the matter system must have a central charge
independent of $\cF$. This is indeed so since 
the normal-ordered Virasoro generators are 
\be
L_k=\frac{1}{4\a'}
: \sum_{n\in\Z}\left(M_{ij}a^i_{k-n}a^j_{n}+a^a_{k-n}a^a_n\right) :
\, 
\ee
and they satisfy the standard Virasoro algebra \cite{CH}
\be
[L_m, L_n] = (m-n) L_{m+n} +\frac{d}{12} m(m^2-1) \d_{m+n}, \quad
d =\, \mbox{spacetime dimension},
\ee
with a central charge unmodified by $\cF$.

\section{Charged String}

We now come to the case of of a charged
open string in background fields or an open string ending 
on two different D-branes with different worldvolume field strengths.
Most part of the analysis 
has already appeared in \cite{charged}. We will go
over some of the salient features. 
The method we used in sec. 2 can be easily applied here. 
Consider an open string with charges $q_1$ and $q_2$ at the
endpoints, the action is ($\a'=1/2$)
\be
S= \frac{1}{2\pi} \int d\tau d\s (\dot{X}_\mu \dot{X}^\mu -
X_\mu' X^{\mu'} )- 
\frac{1}{\pi} \int d \tau (q_1 A_i \dot{X}^i(\s=0) +q_2 A_i
\dot{X}^i(\s=\pi))
\ee
and the boundary conditions are
\bea
&X^{i'}= q_1 F^i{}_j \dot{X}^j, \quad \s=0, \\
&X^{i'}= -q_2 F^i{}_j \dot{X}^j, \quad \s=\pi.
\eea
We will concentrate on a $2 \times 2$ block of $F$,
\be
F= \pmatrix{0 &f \cr -f &0} . 
\ee
Introducing $X_{\pm} = \frac{1}{\sqrt{2}} (X_1 \pm i X_2)$,
the boundary conditions are diagonalized
\be \label{bc}
X_+' = -i \a \dot{X}_+, \quad \s=0;\quad\quad
X_+' = i \b \dot{X}_+, \quad \s=\pi,
\ee
with $\a= q_1 f, \b =q_2 f$. 
In the gauge $A_i = -\frac{1}{2} F_{ij} X^j $, the conjugated momentum are
\bea
&\pi P_- = \dot{X}_+ - \frac{i}{2} X_+ [\a \d(\s) + \b \d( \pi - \s)],
\\
&\pi P_+ = \dot{X}_- + \frac{i}{2} X_- [\a \d(\s) + \b \d( \pi - \s)].
\eea
The same argument as in the previous section shows that the standard
canonical commutation relations have to be modified at the boundary
due to the boundary conditions. 

One can expands $X_\pm$ as  
\bea
&X_+ = x_+ + i \sum_{n > 0} a_n \psi_n - i \sum_{m \ge 0} b_m^\dag
\psi_{-m},\label{X+mode} \\
&X_- = x_- + i \sum_{m \ge 0} b_m \bar{\psi}_{-m} - i \sum_{n > 0}
a_n^\dag  \bar{\psi}_n, \nn
\eea
where we have taken into account $X_-^\dagger= X_+$ and $x_+ =
x_-^\dag$ and the normalized  mode functions for any integer $n$ 
is given by 
\be
\psi_n = \frac{1}{|n-\e|^{1/2}} \cos [(n-\e)\s + \c] e^{-i (n-\e) \t} 
\ee
with $\e = \frac{1}{\pi} (\c+\c')$ and  $\c= \tan^{-1} \a$, $\c'=
\tan^{-1} \a'$.    
$\psi_n$ and the constant mode form a complete basis.
$\psi_n$'s satisfies the boundary condition \eq{bc} and the orthogonality
condition
\be \label{ortho}
\langle \psi_m, \psi_n \rangle = \d_{mn} \mbox{sign}(m-\e), \quad
\langle 1, \psi_n \rangle = 0,
\ee
where the inner product  is defined by 
\be 
\langle f,g \rangle = \frac{1}{\pi} \int_0^\pi d \s \bar{f}(\t,\s)
[ i \stackrel{\leftrightarrow}{\del_\t} + \a \d(\s) + 
\b \d( \pi - \s) ] g(\t,\s) .   
\ee

The symplectic form is given by
\be
\Omega = \int d\s \; (\bfd P_+ \bfd X_+ + \bfd P_- \bfd X_-)
\ee
and it is straightforward to show that it is time independent. 
In terms of $\langle , \rangle$, $\Omega$ can be written compactly as
\be
\Omega =  i \langle \bfd X_+ , \bfd X_+ \rangle.
\ee
Using \eq{ortho}, it is then easy to get
\bea
-i \Omega =   \bfd x_- \bfd x_+ \frac{\a+\b}{\pi}
+\sum_{n >0} \bfd a_n^\dag \bfd a_n   + 
\sum_{m \ge0}\bfd b_m^\dag \bfd b_m  . 
\eea
This implies the nonvanishing commutation relations
\bea 
&[x_+, x_-] = \frac{\pi}{\a+\b},\label{modecr1} \\
&[a_k, a_n^\dag] = \d_{nk}, \quad k,n >0, \\
&[b_l,b_m^\dag] = \d_{lm}, \quad l,m \ge 0,\label{modecr3}
\eea
with all the other commutators zero. 
These relations  
are exactly those obtained in \cite{charged}. However, as we
will see now,  the commutation relation for $[X_+(\t,\s),
X_-(\t,\s')]$ are different. 
Substituting \eq{modecr1}-\eq{modecr3}
back into \eq{X+mode}, one finds
\be
[X_+(\t,\s), X_-(\t,\s')] = J(\s,\s'),
\ee
where
\be
J(\s,\s') = 
 \frac{\pi}{\a+\b} +\sum_{n=-\infty}^{\infty} \frac{1}{n-\e} 
\cos((n-\e)\s+ \c) \cos((n-\e)\s'+ \c). \label{j2}
\ee
We first  compute $J(\s,\s')$ for $\s=\s'=0$ or $\pi$. Using the
identity 
\be
\sum_{n= 1}^{\infty} \frac{2\e}{\e^2 -n^2} + \frac{1}{\e} = \pi \cot
\pi \e, \quad \e \neq \mbox{integer},
\ee
one obtains $J(0,0)= \pi \a/(1+\a^2)$, 
$J(\pi,\pi)= \pi \b /(1+\b^2)$. 
As for $J(\s, \s') $ for other values of $\s,\s'$, it is not hard to
show  that it is zero.  To see this, 
we first remark that it is straightforward 
to show that  $\frac{\del}{\del \s}J 
=0$ for $\s, \s'$ not both $0$ or $\pi$. Therefore $J$ is a constant
for this range of $\s, \s'$ and hence it is sufficient to
calculate $J(0,\pi)$. The latter is equal to
\be
J(0,\pi) = \frac{\pi}{\a+\b} +\cos \c \cos \c'
\sum_{n=-\infty}^{\infty} \frac{(-1)^n}{n-\e}.
\ee
Now it is known that \cite{cpx} 
for any meromorphic function $h(z)$ with poles
$a_1, \cdots, a_m$ (not integers) and with $z= \infty$ a zero of order
$p \geq 2$, the following holds 
\be
\lim_{N\rightarrow \infty} \sum_{n=-N}^{N} 
(-1)^n h(n) = - \pi \sum_{k= 1}^m
\mbox{Res}(\frac{h(z)}{\sin \pi z}, a_k).
\ee
Applying this  for $h= 1/(z^2 -\e^2)$, we get $J(0,\pi)=0$. 
Therefore we obtain finally
\be \label{ch}
[X_+(\t,\s),X_-(\t,\s')] = \left\{
\begin{array}{lll}
\frac{\pi \a}{1+\a^2} , & \s=\s' =0,\cr
\frac{\pi \b}{1+\b^2} , & \s=\s' =\pi,\cr
0, &  \mbox{otherwise},  
\end{array}
\right. 
\ee
The geometrical meaning is clear: 
the noncommutativity is localized at the endpoints and is
determined by the field strength there. If we consider an open string
ending on two different D-branes, the same results \eq{ch} are
obtained; and they agree with the commutation relations \eq{F2}
obtained by quantizing individual
open string with both ends ended on each same D-brane. This is
consistent with the fact that the noncommutativity is a property of
the D-brane and does not depend on what probe we use to see the
noncommutativity. 

The results \eq{ch} can also be obtained by carrying out the Dirac
constrained quantization with the boundary conditions treated as 
constraints \cite{CH,CH1}. The procedure was carried out at the level
of  fields in \cite{CH1}. We mention that one may also express the
boundary conditions as constraints  on the modes 
and carry out the constrained quantization \cite{lee}. 

One may try to recover these results using 
the approach of worldsheet perturbation \cite{Schom}, but it is
difficult to proceed since now the
perturbation due to the charges cannot be written as a worldsheet
action unless one is willing to use delta function. It is also
difficult to proceed  in the line of \cite{SW} as
the Green function  
that satisfies different BC at the two end points is not available.
The advantage of the present approach is apparent, both the 
neutral and charged string can be treated uniformly. 

Two interesting applications of the charged string system
are the studies of creation of open string in electric field and 
D-brane scattering \cite{bachas}. Notice that in these calculations,
only the relations \eq{modecr1}-\eq{modecr3} are used, 
but not \eq{ch}. 

\section{Outlooks}

The kind of noncommutativity that appears in string theory
so far are on the
worldvolume of brane and can be called
``worldvolume'' noncommutativity. Physically this kind 
of noncommutativity arises from the open string interaction and 
has nothing to do with gravity. This is to be contrasted with
another kind of noncommutativity that is due to quantum 
gravity effects at small distance scale.  
We refer to this as ``spacetime'' noncommutativity. It is often
believed that at the Planck scale, spacetime will become fuzzy since
the quantum fluctuation of the geometry cannot be ignored
anymore. 
Since string theory provide a consistent treatment of quantum
gravity, it would be very interesting to understand this better 
from within string theory. 
Consider a D-string with a NS-NS flux along it.  Doing a S-duality  
turns the NS-NS flux into a RR flux and  the original 
noncommutative D-string into a fundamental string with a 
noncommutative worldsheet \cite{CHK}. 
It seems RR background is likely to be relevant for 
``spacetime'' noncommutativity. String theory in RR background
is notoriously difficult, see \cite{nonsp1,nonsp2} 
however for some proposals. 


\section*{Acknowledgments}

I would like to thank Bogdan Morariu and Bruno Zumino for discussions and 
Pei-Ming Ho for  discussions and collaborations.
This work was partially supported by the Swiss National Science
Foundation, by the European Union under TMR contract
ERBFMRX-CT96-0045 and by the Swiss Office for Education and
Science.


\end{document}